\documentclass[]{raa}            
\usepackage{graphicx,times}
\usepackage{natbib}

\providecommand{\bm}[1]{\mbox{\boldmath$#1$\unboldmath}}

\begin{document}

   \title{Comment on ``Constraining the smoothness parameter and dark energy using observational H(z) data''
}

 \volnopage{ {\bf 2011} Vol.\ {\bf X} No. {\bf XX}, 000--000}
   \setcounter{page}{1}

   \author{V. C. Busti  
      \inst{1}
   \and R. C. Santos
      \inst{1,2}
      }

\institute{Instituto de Astronomia, Geof\'isica e Ci\^encias
Atmosf\'ericas, Universidade de S\~ao Paulo, S\~ao Paulo, 05508-090 SP, Brazil; {\it vcbusti@astro.iag.usp.br}
  \and 
  Departamento de Ci\^encias Exatas e da Terra, Universidade
Federal
de S\~ao Paulo (UNIFESP), Diadema, 09972-270  SP, Brazil; {\it cliviars@astro.iag.usp.br} 
}
\vs \no
   {\small Received [year] [month] [day]; accepted [year] [month] [day] }

\abstract{In this Comment we discuss a recent analysis by Yu et al.
[RAA 11, 125 (2011)] about constraints on the smoothness $\alpha$
parameter and dark energy models using observational $H(z)$ data. It
is argued here that  their procedure is conceptually inconsistent
with the basic assumptions underlying the adopted Dyer-Roeder
approach. In order to properly quantify the influence of the $H(z)$
data on the smoothness $\alpha$ parameter, a $\chi^2$-test involving
a sample of SNe Ia and $H(z)$ data in the context of a flat
$\Lambda$CDM model is reanalyzed. This result is confronted  with an
earlier approach  discussed by Santos et al. (2008) without $H(z)$
data. In the ($\Omega_m, \alpha$) plane, it is found that such
parameters are now restricted on the intervals $0.66 \leq \alpha
\leq 1.0$ and $0.27 \leq \Omega_m \leq 0.37$ within 95.4\%
confidence level (2$\sigma$), and, therefore, fully compatible with
the homogeneous case. The basic conclusion is that a joint analysis
involving $H(z)$ data can indirectly improve our knowledge about the
influence of the inhomogeneities. However, this happens only because
the $H(z)$ data provide tighter constraints on the matter density
parameter $\Omega_m$.  }

   \authorrunning{V. C. Busti \& R. C. Santos}            
   \titlerunning{Comment on \cite{yu}}  
   \maketitle

It is widely known that the Universe is homogeneous and isotropic
only at very large scales $(\ga100\,Mpc)$.  In  moderate and smaller
scales, the Universe is inhomogeneous. Since the light propagation
probes the local gravitational field, the clumpiness of matter may
affect the determination of physical parameters comparatively to the
standard Friedmann-Robertson-Walker (FRW) geometry.
\cite{zeldovich1964} and \cite{kantowski1969} were the first to
study this kind of effect. Later on,
\cite{dyerroeder1972,dyerroeder1973} introduced the smoothness
parameter $\alpha$ to quantify the effect of the inhomogeneities in
the magnification of a light beam. For $\alpha=0$ (empty beam), all
the matter is clumped and for $\alpha=1$ the homogeneous case is
recovered. Therefore, the smoothness parameter is restricted over
the interval $[0,1]$. For a clumpy Universe ($\alpha \neq 1$), a new
distance is derived  which is sometimes called the Dyer-Roeder
distance \citep{SEF1992}.

Efforts to obtain observational bounds over $\alpha$ were initially
based on supernovae type Ia (SNe Ia) \citep{santos2008a} and compact
radio sources (Alcaniz, Lima \& Silva 2004; Santos \& Lima 2008). In particular, by 
assuming that the dark energy is a smooth component, Santos et al. (2008) obtained
$\alpha \geq 0.42$ within 95.4\% confidence level with basis on the
\cite{riess} SNe Ia sample. It was  also shown that compact radio
sources \citep{Gurvits1999,LA2000,LA2002} did not constrain $\alpha$
\citep{ALS2004,santos2008b}.

Recently, \cite{yu} claimed that better constraints over the
Dyer-Roeder parameter $\alpha$ could be obtained based only on the
observational $H(z)$ data. By using a $\chi^2$ minimization method
they found $\alpha=0.81^{+0.19}_{-0.20}$ and $\Omega_{\rm
M}=0.32^{+0.12}_{-0.06}$ at $1\sigma$ confidence level. Further, by
assuming a Gaussian prior of $\Omega_{\rm M}=0.26\pm0.1$, the limits
$\alpha=0.93^{+0.07}_{-0.19}$ and $\Omega_{\rm
M}=0.31^{+0.06}_{-0.05}$ were also derived. Finally, for a XCDM model,
the smoothness parameter was constrained to $\alpha\geq0.80$ with
$\omega$ weakly constrained around -1, where $\omega$ describes the
equation of the state of the dark energy ($p_{\rm X}=\omega\rho_{\rm
X}$). However, as it will be argued in the present comment, there is
a profound contradiction between their implementation of the
observational Hubble data and the underlying assumptions of the
Dyer-Roeder approach. In other words, the $H(z)$ data alone 
cannot constrain the $\alpha$ parameter. 

To begin with, let us first discuss the basic assumptions of the
Dyer-Roeder procedure. The main hypothesis is that the Universe is
locally inhomogeneous,  where underdensities in voids are
compensated by overdensities in clumps  thereby making the Universe
homogeneous at very large scales. A typical line of sight is far from the
clumps, not suffering from gravitational lensing effects,  being
reasonable to consider that the light beam experiences an effective
$\alpha \rho_m$ matter energy density and negligible shear. On the
other hand, the dynamics is expected to feel the influence of a
volume smoothed description
\citep{bildhauer1991,buchert1997,linder1998} and it is the same as
the homogeneous case. Thus, in the Dyer-Roeder approach, the Hubble
parameter does not depend on the smoothness parameter. Actually, the
homogeneous Hubble parameter is used in the derivation of the
Dyer-Roeder equation, as can be seen in the following differential
equation for the angular diameter distance (see, for instance,
Mattsson, 2010)

\begin{equation}
H(z)\frac{d}{dz}\left[ (1+z)^2 H(z) \frac{d}{dz} d_A (z) \right] +
4\pi G [\rho(z) + p(z)]d_A (z) = 0. \label{eq1}
\end{equation}
In the above expression, $H(z)$ stands for the Hubble parameter,
$d_A (z)$ for the angular diameter distance, $\rho (z)$ for the
total energy density and $p(z)$ for the total pressure. The
smoothness parameter enters only in the second term through the
effective $\rho(z)$ function.

On the other hand, in order to implement the observational Hubble
data, \cite{yu} also  deduced the correspondence between $H(z)$ and
$d_A(z)$ (see their Eqs. (22)-(25))

\begin{equation}
 \frac{H(z)}{H_0}=\frac{1}{(1+z)d^{\prime}_{A}(z)+d_A(z)},
\label{eq2}
\end{equation}
where $H_0$ is the Hubble constant and the prime denotes
differentiation with respect to $z$. However, as one may check, the
above relation is valid only when $\alpha=1$, that is, in the
homogeneous case. In this way, it does not make sense to use a form
for $H(z)$ independent of $\alpha$ to obtain the Dyer-Roeder
distance (see Eq. \ref{eq1}),  and, simultaneously, to consider 
$H(z)$ with a dependence on $\alpha$ as given by the above  equation \ref{eq2}. It thus follows that  the
analysis made by \cite{yu} is both conceptually and mathematically
flawed and, as such, their corresponding results are meaningless.

Nevertheless, the question posed by \cite{yu} concerning a possible
influence of the $H(z)$ data on the constraints of $\alpha$ can still be considered at least in 
the context of a joint analysis, for instance, 
involving supernovas and other cosmological tests. 
In this case, since the observational Hubble
data constrain the cosmology itself, it is interesting to quantify how
the $H(z)$ data can modify the limits established on the smoothness parameter using, for instance, 
the SNe type Ia analysis appearing in the paper by \cite{santos2008a}.  Naturally, one may think that
the effect must be small because the  $H(z)$ does not depend explicitly on the
value of $\alpha$.

In figure 1a we display the results obtained by \cite{santos2008a}
through a $\chi^2$-test by using only the 182 SNe Ia from
\cite{riess} (gold sample).  In their analysis they obtained  $0.42 \leq  \alpha
\leq 1.0$ and $0.25 \leq \Omega_m \leq 0.44$ at $2\sigma$
confidence level. The corresponding best fits are $\alpha = 1$ and $\Omega_m=0.33$ and, therefore, 
fully in agreement with the homogeneous case. 

In figure 1b,  we show the present joint analysis by considering the
same gold sample plus 12 $H(z)$ data \citep{simon2005,daly2008}. The
parameters are now restricted over the intervals $0.66 \leq \alpha
\leq 1.0$ and $0.27 \leq \Omega_m \leq 0.37$ within $2\sigma$
confidence level. The best fits are $\alpha=1.0$ and $\Omega_m=0.32$.
As should be physically expected, the constraints are mildly
improved by introducing the $H(z)$ data. In particular, the best fit value of the smoothness parameter is given by the homogeneous case ($\alpha=1)$, as derived earlier by Santos et al. (2008).  This fact can be understood by realizing that only the value of $\Omega_m$ is directly constrained
by the $H(z)$ data. Naturally, the present results also suggest that
the remaining analysis for XCDM models studied by \cite{yu} should also 
be rediscussed.

\begin{figure}
\hspace{0.5cm}
\includegraphics[width=60mm]{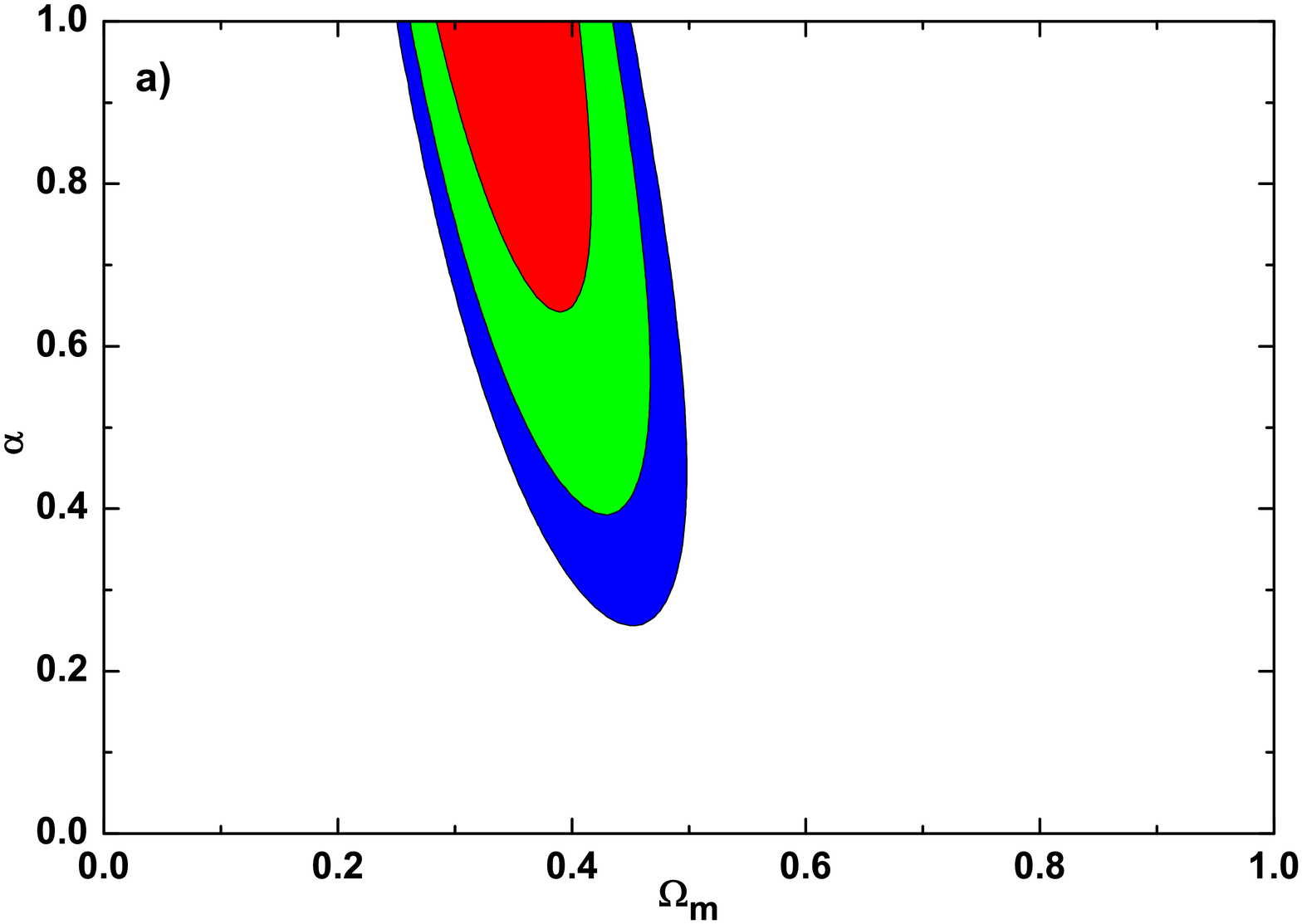} \hspace{0.5cm}
\includegraphics[width=60mm]{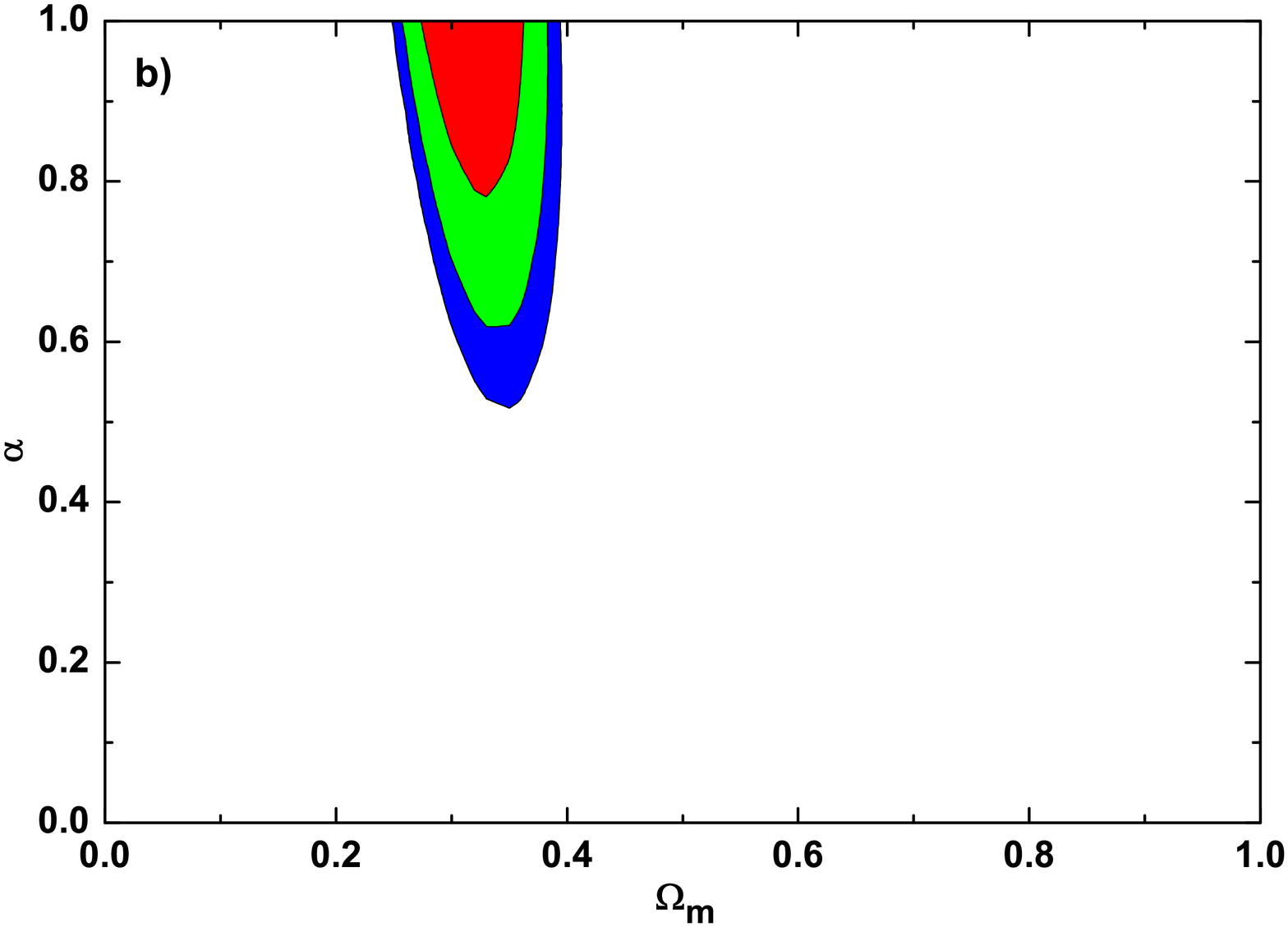}

\caption{{\bf a)} The $\alpha-\Omega_m$ plane for 182 SNe Ia from
\cite{riess} as discussed by Santos et al. (2008). {\bf b)}
Constraints for a joint analysis involving the same SNe Ia sample
plus 12 $H(z)$ data from \cite{simon2005} and \cite{daly2008}. The constraints
obtained with the joint analysis are $0.66 \leq \alpha \leq 1.0$ and
$0.27 \leq \Omega_m \leq 0.37$ (2$\sigma$), with a best fit of
$\alpha=1.0$ and $\Omega_m=0.32$ (see comments in the text).}
   \label{Fig1}
\end{figure}

\normalem
\begin{acknowledgements}
The authors would like to thank  J. A. S. Lima and J. V.
Cunha for helpful discussions and Tong-Jie Zhang and Hao-Ran Yu for useful correspondence. VCB is supported by CNPq (Brazilian
Research Agency).
\end{acknowledgements}

\label{lastpage}

\end{document}